\documentclass[10pt]{book}
\usepackage{array}
\usepackage{amsthm,amstext,amsmath,amssymb}
\usepackage{graphicx}
\usepackage{bbm}
\usepackage{epstopdf}

\newcommand{\indicator}[1]{\mathbbm{1}_{\{#1\}}}   
\newcommand{\Exp}{\mathbb{E}}
\newcommand{\Pro}{\mathbb{P}}

\newcommand{\cF}{\mathcal{F}}
\newcommand{\cJ}{\mathcal{J}}
\newcommand{\cL}{\mathcal{L}}
\newcommand{\cS}{\mathcal{S}}
\newcommand{\cT}{\mathcal{T}}
\newcommand{\cN}{\mathcal{N}}

\newcommand{\bphi}{\boldsymbol{\phi}}
\newcommand{\bdelta}{\boldsymbol{\delta}}
\newcommand{\brho}{\boldsymbol{\rho}}

\DeclareMathOperator{\gaussdf}{\mathcal{N}}

\DeclareMathOperator*{\esssup}{ess\,sup}

\newcommand{\set}[1]{\left\{#1\right\}}

\begin{document}
\renewcommand{\baselinestretch}{1.2}
\markright{
}
\markboth{\hfill{\footnotesize\rm G.V. MOUSTAKIDES, A.S. POLUNCHENKO AND A.G. TARTAKOVSKY}\hfill}
{\hfill {\footnotesize\rm NUMERICAL ANALYSIS OF CHANGE-POINT DETECTION PROCEDURES} \hfill}
\renewcommand{\thefootnote}{}
$\ $\par
\fontsize{10.95}{14pt plus.8pt minus .6pt}\selectfont
\vspace{0.8pc}
\centerline{\large\bf A NUMERICAL APPROACH TO PERFORMANCE ANALYSIS OF}
\vspace{2pt}
\centerline{\large\bf QUICKEST CHANGE-POINT DETECTION PROCEDURES}
\vspace{.4cm}
\centerline{George V. Moustakides, Aleksey S. Polunchenko and Alexander G. Tartakovsky}
\vspace{.4cm}
\centerline{\it University of Patras and University of Southern California}
\vspace{.55cm}
\fontsize{9}{11.5pt plus.8pt minus .6pt}\selectfont

\begin{quotation}
\noindent {\it Abstract:}
For the most popular sequential change detection rules such as CUSUM, EWMA, and
the Shiryaev-Roberts test, we develop integral equations and a concise numerical
method to compute a number of performance metrics, including average
detection delay and average time to false alarm. We pay special
attention to the Shiryaev-Roberts procedure and evaluate its performance for
various initialization strategies. Regarding the randomized initialization variant
proposed by Pollak, known to be asymptotically optimal of order-3, we
offer a means for numerically computing the quasi-stationary
distribution of the Shiryaev-Roberts statistic that is the distribution of the initializing random variable, thus
making this test applicable in practice. A significant side-product of our
computational technique is the observation that deterministic initializations of
the Shiryaev-Roberts procedure can also enjoy the same order-3 optimality property
as Pollak's randomized test and, after careful selection, even uniformly outperform
it.\par

\vspace{9pt}
\par
\end{quotation}\par

\fontsize{10.95}{14pt plus.8pt minus .6pt}\selectfont
\setcounter{chapter}{1}
\setcounter{equation}{0} 
\noindent {\bf 1. Introduction}
\\
Change-point problems deal with anomaly detection or more generally detection of
changes in the statistical behavior of processes. This problem has an enormous spectrum of
important applications, including biomedical signal and image processing, quality
control engineering, financial markets, link failure detection in communication
networks, intrusion detection in computer networks and security systems, detection
and tracking of covert hostile activities, chemical or biological warfare agent
detection systems (as a protection tool against terrorist attacks), detection of
the onset of an epidemic, failure detection in manufacturing systems and large
machines, target detection in surveillance systems, econometrics, seismology,
navigation, speech segmentation, and the analysis of historical texts. See, e.g.,
Willsky (1976), Basseville and Nikiforov (1993), MacNeill (1993), Baron (2002),
Galstyan, Mitra, and Cohen (2007), Kent (2000), Tartakovsky (1991), Tartakovsky and
Ivanova (1992), Tartakovsky and Veeravalli (2004), Tartakovsky, Li, and Yaralov
(2003), Tartakovsky, Rozovskii, Bla\v{z}ek, and Kim (2006), Wang, Zhang, and Shin
(2002). In all of these applications, sensors monitoring the environment take
observations that undergo a change in distribution in response to changes and
anomalies in the environment or changes in patterns of a certain behavior. The
observations are obtained sequentially and, as long as their behavior is consistent
with the normal state, one is content to let the process continue. If the state
changes, then one is interested in detecting the change as soon as possible while
minimizing false detections.

Let $\{X_n\}_{n\ge 1}$ denote observations that are obtained
sequentially, and let $\Pro_\infty$ and $\Pro_0$ be the probability measures
before and after the change. Let $\Pro_\tau$ and $\Exp_\tau$ denote the
probability measure and the expectation induced when the time of
change is $\tau\ge0$. We use the convention that $\tau$ is the {\it last}
time instant where the observations follow the nominal (pre-change) regime, 
so the first observation under the alternative measure is at
time $\tau+1$. Thus $\Pro_0$ means that the change took place
before any observations were taken and all observations are under the
alternative regime, whereas $\Pro_\infty$ stands for the scenario in which the change
is at infinity (i.e., does not occur) and all observations are under
the nominal regime.

A sequential change detection procedure is identified with a stopping time $T$
that is adapted to the filtration $\{\cF_n\}_{n\ge0}$, where $\cF_0$ is the
trivial $\sigma$-algebra and, for $n\ge 1$, $\cF_n=\sigma\{X_1,\ldots,X_n\}$ is the
$\sigma$-algebra generated by the first $n$ observations. Thus the event
$\{T\le n\}$ belongs to $\cF_n$. Since the change occurs at an unknown instant, the
objective is to detect it as quickly as possible while avoiding frequent false
alarms. Therefore, the
design of the quickest change-point detection procedure involves optimizing the
trade-off between two kinds of performance measures, one being a measure of
detection delay and the other being a measure of the frequency of false alarms.
Regarding the latter, the false alarm rate is usually measured by the average time
to false alarm $\Exp_\infty[T]$, commonly referred to as the average run
length (ARL) to false alarm. For detection delay, two major non-Bayesian
criteria have been proposed in the literature. The first, due to Lorden (1971), is
\begin{align}\label{eq:lorden}
\cJ_{\rm L}(T)=\sup_{\tau\ge0}\esssup\Exp_\tau[(T-\tau)^+|\cF_\tau],
\end{align}
where $x^+=\max\{x,0\}$; and the second, due to Pollak (1985), is
\begin{align}\label{eq:pollak}
\cJ_{\rm P}(T)=\sup_{\tau\ge0}\Exp_\tau[T-\tau|T>\tau],
\end{align}
where $\Exp_\tau[T-\tau|T>\tau]$ is the (conditional) average delay to detection
for the fixed point of change $0 \le \tau <\infty$. As discussed in
Moustakides (2008), Lorden's performance measure is appropriate for problems in
which the change-point mechanism takes into account the observations for deciding
about imposing the change. Pollak's measure, on the other hand, assumes that the
change is imposed by a source independent from the observations. Both cases are
equally important, referring to completely different classes of change-point
applications.

Whether we use Lorden's or Pollak's measure, in finding an optimal stopping time one would normally be interested in minimizing $\cJ_{\rm L}(T)$ or $\cJ_{\rm
P}(T)$ and maximizing, at the same time, $\Exp_\infty[T]$. The
two goals are antagonistic and, therefore, we are content to minimize the worst
average detection delay while controlling the ARL to false alarm $\Exp_\infty[T]$
above a prescribed level. More formally, we are interested in solving the minimax constrained optimization problems
\begin{equation}
\inf_T\cJ_{\rm L}(T)=
\inf_T\sup_{\tau\ge0}\esssup\Exp_\tau[(T-\tau)^+|\cF_\tau];~\mbox{\rm subject
to}~\Exp_\infty[T]\ge\gamma \label{eq:problemL}
\end{equation}
for Lorden's measure, or
\begin{equation}
\inf_T\cJ_{\rm P}(T)=\inf_T\sup_{\tau\ge0}\Exp_\tau[T-\tau|T>\tau];~\mbox{\rm
subject to}~\Exp_\infty[T]\ge\gamma \label{eq:problemP}
\end{equation}
for Pollak's measure. In both cases $\gamma\ge1$ is the prescribed minimum value of
the ARL to false alarm. The problems~\eqref{eq:problemL} and~\eqref{eq:problemP} are central to sequential change-point detection theory, and numerous past and ongoing efforts aim to find the corresponding solutions
for various observation models.

Regarding existing optimality results, Lorden (1971) proved that the Cumulative Sum
(CUSUM) procedure, introduced by Page (1954), asymptotically (as
$\gamma\to\infty$) solves the minimax constrained optimization problem in~\eqref{eq:problemL} for i.i.d. observations before and after the change. Later,
Moustakides (1986) showed that CUSUM is {\em exactly} optimal for every $\gamma>1$
(see also Ritov (1990)). An analogous result for detecting a change in the drift of
a Brownian motion has been independently established by Beibel (1996) and Shiryaev
(1996). For an extension to a Gaussian process with independent nonhomogeneous
increments see Tartakovsky (1995), and for a generalization to It\^o
processes see Moustakides (2004).

Shiryaev (1961) introduced an alternative to the CUSUM change detection scheme,
currently known as the Shiryaev-Roberts (SR) procedure (cf.~Roberts (1966)), that is central here.
A randomized variant of
this test was proposed by Pollak (1985) where, instead of initializing the test
from 0 as in the original version, Pollak suggested a randomized initialization
strategy with the initial point sampled from the quasi-stationary distribution of the SR statistic. We
refer to this version as the Shiryaev-Roberts-Pollak (SRP) procedure. The gain
obtained by this alternative initialization mechanism is significant. Pollak
(1985) was able to demonstrate that his variant solves, asymptotically as
$\gamma\to\infty$, the optimization problem defined in~\eqref{eq:problemP} within
an $o(1)$ quantity. More precisely, the SRP procedure has a $\cJ_{\rm P}$ measure
that differs from the (unknown) optimum by a quantity that tends to 0 as
$\gamma\to\infty$, even though both worst average detection delays {\em tend to
infinity}. We refer to this asymptotic optimality as {\em order-3}, as
opposed to order-1, when the ratio of the two quantities tends to 1, or order-2 when
their difference is bounded.

Despite its strong asymptotic optimality property, the SRP procedure
is impossible to apply in practice because there is neither an analytical nor a numerical method for computation of the quasi-stationary distribution required for the initializing random variable  (except in some rare cases; see, e.g., Pollak (1985) and  Mevorach and Pollak (1991)). An important result of the present paper is a solid numerical technique for the computation of this distribution, making the SRP procedure readily available for applications. In addition, we examine alternative {\em deterministic} initialization strategies for the SR test. These variants, as we shall see in our numerical examples, are strong competitors of SRP, enjoying the same order-3 asymptotic optimality. As a matter of fact, in all of the examples we have tried, our tests (with several proposed initialization schemes) either outperformed the SRP test, or performed equally well in the sense that they exhibited either smaller or the same $\cJ_{\rm P}$ measure for the same ARL to false alarm. At the same time, the proposed versions of the SR procedure have a {\em fast initial response}, providing a smaller average detection delay for changes that occur from the very beginning (and soon after surveillance begins) as compared to both the conventional SR and the SRP procedure.

Regarding the classical SR procedure, Shiryaev (1961, 1963) considered the problem of detecting a change in the
mean of a Brownian motion when a stationary regime is in place, effected by a
change possibly occurring in a distant future, after many false alarms have been
experienced. Shiryaev proved that the SR procedure is {\em exactly} optimal for
minimizing the expected delay in detecting such distant changes against a
stationary background of false alarms. Recently Pollak and Tartakovsky (2009),
motivated by this result and by the work of Feinberg and Shiryaev (2006), obtained
a similar result for detecting a change in a general discrete-time model, assuming
that a change occurs at a far horizon (i.e., when $\tau$ is large) and is preceded
by a stationary flow of false alarms. Specifically, the SR procedure was shown to
be {\em exactly} optimal in minimizing, subject to the familiar constraint
$\Exp_\infty[T]\ge\gamma$, the {\em relative integral average detection delay}
$$
\cJ(T)=\frac{\sum_{\tau=0}^\infty\Exp_\tau[(T-\tau)^+]}{\Exp_\infty[T]}
$$
instead of the worst expected conditional detection delay $\cJ_{\rm P}(T)$ of~\eqref{eq:pollak}. 
Furthermore, for a general discrete-time model, the value of
$\cJ(T)$ has been shown to be equal to the limiting (as $\tau\to\infty$) value of
the average detection delay of the repeated SR detection procedure when the same
stopping time is reapplied after each false alarm. This result was
initially established by Shiryaev (1961, 1963) for the Brownian motion model.

Finding the appropriate version of the SR procedure that minimizes Pollak's
$\cJ_{\rm P}(T)$ measure is still an open problem. Answering this question is
essential because the corresponding optimal test constitutes the missing complement
of the CUSUM procedure for the two drastically different classes of change detection applications
mentioned earlier.

Concluding the literature review on the CUSUM and SR tests, Tartakovsky and Ivanova (1992) consider the general case of processes with independent increments (for discrete and continuous time), providing
efficient asymptotic formulas for the performance of the two procedures. Earlier,
Pollak and Siegmund (1985) carried out a similar analysis for the Brownian motion
case.

A third test is the exponentially weighted
moving average (EWMA) procedure, first proposed by Roberts (1959). Its behavior was
studied in detail by Novikov and Ergashev (1988) and Novikov (1990) for arbitrary
processes with independent and homogeneous increments. They show that the optimized EWMA procedure exhibits 23\% more expected detection
lag as compared to the CUSUM or SR procedure when detecting a change in the mean of a
Gaussian process. These results were corroborated by Srivastava and Wu
(1993) for detecting a change in the drift of a Brownian motion using an
alternative technique. A comprehensive analysis of various EWMA schemes, and how
they compete with CUSUM, can also be found in Lucas and Saccucci (1990).

The main goal here is a simple numerical method for the
evaluation of the operating characteristics of the SR test and its SRP variant. Our
numerical technique is also used for the performance evaluation and
optimization of an alternative version of the SR procedure in which the
initializing value is deterministic instead of random (which is the case in the SRP
procedure). The final detection procedure that comes out of this optimization
(as well as its modifications based on various initial conditions) is compared
against the SRP test. In all numerical examples we present, the
optimized deterministic initialization enjoys the same order-3 asymptotic
optimality property as the SRP test and, more importantly, uniformly outperforms
it. Of course these claims are only observations based on our numerical findings, but we work to support them analytically. In fact, a proof that the SR test with a certain deterministic initialization is exactly minimax (while the SRP test is not) in a particular example can be found in Polunchenko and Tartakovsky (2010). 

This article is organized as follows. In Section~2 we formally state the problem and outline the SR test and its variants. In Section~3 we develop a system of {\em exact} integral equations on the performance metrics. In the same section we propose a set of approximations to these equations that arise when we develop numerical solutions to the initial set of integral equations. In Section~4 we give numerical examples involving Gaussian and exponential models to illustrate the capabilities of our numerical methodology, and compare the relative performance of the SR test and its variants of interest. Additionally in Subsection~4.3 we show, very briefly, how our computational method can be modified to suit the other two popular tests -- CUSUM and EWMA. Finally, in Section~5 we provide a summary of conclusions that can be drawn from our study.

\null\par

\setcounter{chapter}{2}
\setcounter{equation}{0} 
\noindent {\bf 2. The Shiryaev-Roberts Test and its Variants}
\\
We provide a brief overview of the SR test and its
randomized variant -- the SRP test -- and introduce the version with the deterministic initialization that we propose here as an alternative to the SRP test.

We make the following assumptions regarding the change-point detection problem. Suppose a sequence~$\set{X_n}_{n\ge 1}$ of i.i.d.~random
variables is observed sequentially. Initially the sequence is ``in-control'', i.e.,
all the observations come from pdf $f_\infty(x)$. At an unknown time
$\tau\ge0$, something happens and the sequence runs ``out of control'' by
abruptly changing its statistical properties, so that from $\tau+1$ on the pdf switches to $f_0(x)\not\equiv
f_\infty(x)$. At this point it is desired to raise an alarm as quickly as possible,
allowing for an appropriate action to be taken. We recall that a sequential
detection procedure is identified with a stopping time $T$ that is adapted
to the filtration $\set{\cF_n}_{n\ge0}$ generated by the observations.

To define the SR procedure let $\ell_n=f_0(X_n)/f_\infty(X_n)$ denote the
likelihood ratio of the $n$th observation and let $R_n$ be the SR statistic defined
as
\begin{equation}
R_n=\sum_{k=1}^n\prod_{j=k}^n \ell_j. \label{eq:statistics1}
\end{equation}
Then the original SR stopping time $\cS_\nu$ is the first time $n$ that $R_n$
attains a positive level $\nu$, i.e.,
\begin{equation}
\cS_\nu=\inf\{n\ge1:R_n\ge\nu\},~{\rm with}~\inf\{\varnothing\}=\infty,
\label{eq:st1}
\end{equation}
where threshold $\nu =\nu_\gamma$ is selected so that the false alarm constraint is
satisfied with equality, i.e., $\Exp_\infty[\cS_{\nu_\gamma}]=\gamma$. It is easy
to verify from~\eqref{eq:statistics1} that the SR statistic follows the
recursion $R_n=(1+R_{n-1})\ell_n$ initialized with $R_0=0$.

We propose a modification by initializing the test from any value $R_0=r\ge0$. 
Define the modified SR statistic $R_n^r$ by the recursion
\begin{equation}
R_n^r=\left(1+R_{n-1}^r\right)\ell_n,~~R_0^r=r, \label{eq:statistics2}
\end{equation}
and the corresponding stopping time 
\begin{equation}
\cS^r_\nu=\inf\{n\ge1:R^r_n\ge\nu\},
\label{eq:st2}
\end{equation}
where again $\nu$ is selected so that $\Exp_\infty[\cS_{\nu}^r]=\gamma$. We
call this variant SR-$r$.
Clearly, threshold $\nu$ and initializing value $r$ are related through the
equation $\Exp_\infty[\cS_\nu^r]=\gamma$. In satisfying this equality we can either
assume that $\nu$ is a function $\nu_r$ of $r$, or that $r$ is a
function $r_\nu$ of threshold $\nu$. For simplicity we omit
subscripts.

Our intention is to
isolate a specific value for the initializing parameter $r$ that will give rise to
a test that competes effectively with Pollak's randomized SRP version.

The SRP procedure is defined similarly to~\eqref{eq:statistics2} and~\eqref{eq:st2} but with $R_0$ now a random variable distributed according to the quasi-stationary distribution of the SR statistic $R_n$:
\begin{equation}
\Pro[R_0 \le x]=\lim_{n\to\infty}\Pro_\infty[R_n^0\le x|\cS_\nu^0 > n],
~x\in[0,\nu). \label{eq:quasi}
\end{equation}
To avoid complications we assume that the likelihood ratio
$\ell_1=f_0(X_1)/f_\infty(X_1)$ is continuous, in which case the quasi-stationary
distribution exists (cf.~Harris (1963), Theorem III.10.1). However, the case where
$\ell_1$ is nonarithmetic can also be covered with some additional effort.

Let the quasi-stationary density of the distribution \eqref{eq:quasi} be $q(x)$. The SRP procedure is defined by
\begin{align}
R_n^q&=\left(1+R_{n-1}^q\right)\ell_n,~~R_0^q\sim q(x),\label{eq:srp1}\\
\cS^q_\nu&=\inf\{n\ge1:R^q_n\ge\nu\},
\label{eq:srp2}
\end{align}
where $R_0^q\sim q(x)$ means that the initializing variable is random and
distributed according to the pdf $q(x)$. Note that $q(x)=q_\nu(x)$ depends on $\nu$
and its support is $[0,\nu)$. Again, the threshold $\nu$ is selected so that
$\Exp_\infty[\cS_\nu^q]=\gamma$. The main drawback of this test has been the fact
that there was no specific way to compute $q(x)$, and finding $q(x)$ has been an open
problem since the first appearance of the SRP test in 1985. In Section~3 we give an efficient numerical answer.

To understand the reason for the randomized
initialization, we observe that $\cJ_{\rm P}(T)$ is the supremum over the
change time $\tau$ of the sequence of conditional expected detection delays
$\Exp_\tau[T-\tau|T>\tau],~\tau\ge0$. According to the general decision theory
(see, e.g., Ferguson (1967), Theorem 2.11.3) if a)~we can find a stopping time $T$
{\it adapted to} $\{\cF_n\}_{n\ge0}$ that is
an extended Bayes and an {\em equalizer rule} and b)~$T$ satisfies the
false alarm constraint with equality, then $T$ solves the minimax constrained
optimization problem defined in~\eqref{eq:problemP}. It follows from Pollak
(1985) that the randomized initialization according to the quasi-stationary
distribution guarantees the equalizer property
$\Exp_0[\cS^q_\nu]=\Exp_\tau[\cS^q_\nu-\tau|\cS^q_\nu>\tau]$ for all $\tau\ge0$
and that threshold $\nu=\nu_\gamma$ can be selected
in such a way that the false alarm constraint is satisfied with equality. However,
the SRP test was shown in Pollak (1985) to be only asymptotically optimal of
order-3: if $\nu=\nu_\gamma$ is such that $\Exp_\infty[\cS^q_\nu]
= \gamma$, we have
\begin{equation}\label{AO-SRP}
\Exp_0[\cS^q_\nu]-\inf_{\{T: \Exp_\infty [T] \ge \gamma\}}\sup_{\tau \ge 0}
\Exp_\tau[T-\tau|T>\tau]=o(1)\quad \text{as $\gamma \to \infty$}.
\end{equation}
The question of which test exactly optimizes $\cJ_{\rm P}(T)$ and solves the minimax problem  \eqref{eq:problemP} is still open.

Due to \eqref{AO-SRP}
and the fact that the SRP is an equalizer, one can conjecture that this
test is exactly optimal. No further analysis or
counterexamples were offered until recently to support or disprove this (see also Mei (2006)). We believe that the proposed
SR-$r$ variant can in fact provide a counterexample. As we see from the numerical examples in Section~4, the SR-$r$ test, properly optimized, can perform uniformly better than the SRP test. Although this is only based on our numerical findings, it nevertheless provides a strong evidence against the exact optimality of the SRP procedure. A counterexample where the SRP test is not optimal but the proposed SR-$r$ test is optimal can be found in Polunchenko and Tartakovsky (2010). While we believe that the optimal (for any given $\gamma$) solution is a specially designed (non-randomized) SR-$r$ test with a varying in time (increasing) threshold (to guarantee constant conditional average detection delay), further discussion is out of the scope of this paper and will be presented elsewhere.

The idea of initializing the test statistics with a value different
from 0 has been applied in the past by Lucas and Crosier (1982) and Lucas
(1985) to CUSUM. The goal there was to reduce the average detection delay when
observations are affected by a change from the beginning
or soon after surveillance begins. By means of Monte Carlo simulations, it was
shown that CUSUM with a positive head start exhibits the so-called fast initial
response feature, permitting a more rapid response to an initial
``out-of-control'' situation, than does the conventional CUSUM (initialized from 0),
at a price of a minor performance degradation for large values of the point of
change $\tau$. However, no method for choosing this initial head start
value has been proposed.

\vskip0.4cm\par
\noindent {\bf 2.1 Lower Bound and Asymptotic Optimality of Order-3}
\\
To assess the quality of a detection scheme we
compare the test of interest against a {\it lower bound} of the optimal
performance. Finding such a bound turns out to be much easier than finding the
optimal test.

We now show that the SR-$r$ test with initial condition $R_0^r=r$ can provide a convenient lower bound for the optimal performance. Indeed,
observe that for any stopping time $T$ and any point of change $\tau\ge0$ we can
write
$$
\cJ_{\rm P}(T)\ge\Exp_\tau[T-\tau|T>\tau]=\frac{\Exp_\tau[(T-\tau)^+]}{\Pro_\tau[T>\tau]} = \frac{\Exp_\tau[(T-\tau)^+]}{\Pro_\infty[T>\tau]},
$$
where we used the fact that since at $\tau$ we are still under nominal conditions, we have
$\Pro_\tau[T>\tau]=\Pro_\infty[T>\tau]$. From the previous inequality we conclude
that
\begin{align}
\cJ_{\rm P}(T)\Pro_\infty[T>\tau]\ge\Exp_\tau[(T-\tau)^+].
\label{eq:aux1}
\end{align}
Applying this inequality for $\tau=0$, multiplying each side with $r\ge0$ and
observing that $\Pro_\infty[T>0]=1$, we deduce that
\begin{align}
r\cJ_{\rm P}(T)\ge r\Exp_0[T].
\label{eq:aux2}
\end{align}
Summing each side of~\eqref{eq:aux1} over all $\tau\ge0$ and adding the
corresponding sides of~\eqref{eq:aux2}, we end up with the inequality
$$
\cJ_{\rm P}(T)\left\{r+\sum_{\tau=0}^\infty\Pro_\infty[T>\tau]\right\}
\ge\left\{r\Exp_0[T]+\sum_{\tau=0}^\infty\Exp_\tau[(T-\tau)^+]\right\}
$$
or, equivalently,
\begin{align*}
\cJ_{\rm P}(T)&\ge \frac{r\Exp_0[T]+\sum_{\tau=0}^\infty\Exp_\tau[(T-\tau)^+]}
{r+\sum_{\tau=0}^\infty\Pro_\infty[T>\tau]}\\
&= \frac{r\Exp_0[T]+\sum_{\tau=0}^\infty\Exp_\tau[(T-\tau)^+]}{r+\Exp_\infty[T]}.
\end{align*}
If we call the lower bound
\begin{align}
\cL_{\rm P}(T)=\frac{r\Exp_0[T]+\sum_{\tau=0}^\infty\Exp_\tau[(T-\tau)^+]}
{r+\Exp_\infty[T]} \label{eq:bound}
\end{align}
and optimize each side of the previous inequality over all $T$ that satisfy the false alarm constraint $\Exp_\infty[T]\ge\gamma$, we obtain
\[
 \inf_{\{T: \Exp_\infty[T]\ge \gamma\}}\cJ_{\rm
P}(T)\ge\inf_{\{T: \Exp_\infty[T]\ge \gamma\}}\cL_{\rm P}(T).
\]
Fortunately, the optimization of the lower bound $\cL_{\rm P}(T)$ is
possible and the optimizing stopping time is simply $\cS_\nu^r$, that is,
$\inf_{\{T: \Exp_\infty[T]\ge
\gamma\}}\cL_{\rm P}(T)=\cL_{\rm P}(\cS_\nu^r)$, where
$\nu=\nu_\gamma$ is such that $\Exp_\infty[\cS_\nu^r]=\gamma$. The proof of this
statement for $r=0$ is given in Pollak and Tartakovsky (2009), and for any arbitrary
positive $r$ it can be shown following similar arguments. The interesting
observation is that the inequality is true for {\it any} nonnegative value of $r$.

From our previous arguments we have
\begin{equation} \label{LUB}
\cJ_{\rm P}(\cS_\nu^r)\ge\inf_{\{T:
\Exp_\infty[T]\ge \gamma\}} \cJ_{\rm P}(T)\ge \cL_{\rm P}(\cS_\nu^r).
\end{equation}
This double inequality suggests that if we are interested in verifying whether
$\cS_\nu^r$ is asymptotically optimal of order-3, it is sufficient to show that
\begin{equation}
\lim_{\gamma\to\infty}\left\{\cJ_{\rm P}(\cS_\nu^r)-\cL_{\rm
P}(\cS_\nu^r)\right\}=0.
\label{eq:order3}
\end{equation}

Fix a threshold $\nu>0$ and consider the specific initializing value
\begin{equation}
r_\nu={\rm arg}\inf_{0\le r<\nu}\left\{\cJ_{\rm P}(\cS_\nu^r)-\cL_{\rm
P}(\cS_\nu^r)\right\} \label{eq:opt_r}
\end{equation}
as a candidate for initialization of the SR-$r$ scheme. The resulting stopping time $\cS_\nu^{r_\nu}$ is now a function only of the threshold $\nu$, and the latter is selected so that $\cS_\nu^{r_\nu}$ satisfies the false alarm constraint with equality. This uniquely defines our test, since both
$r=r_\gamma$ and $\nu=\nu_\gamma$ depend only on the false alarm parameter
$\gamma$.

\begin{figure}[ht]
 \centerline{\includegraphics[width=0.49\textwidth]{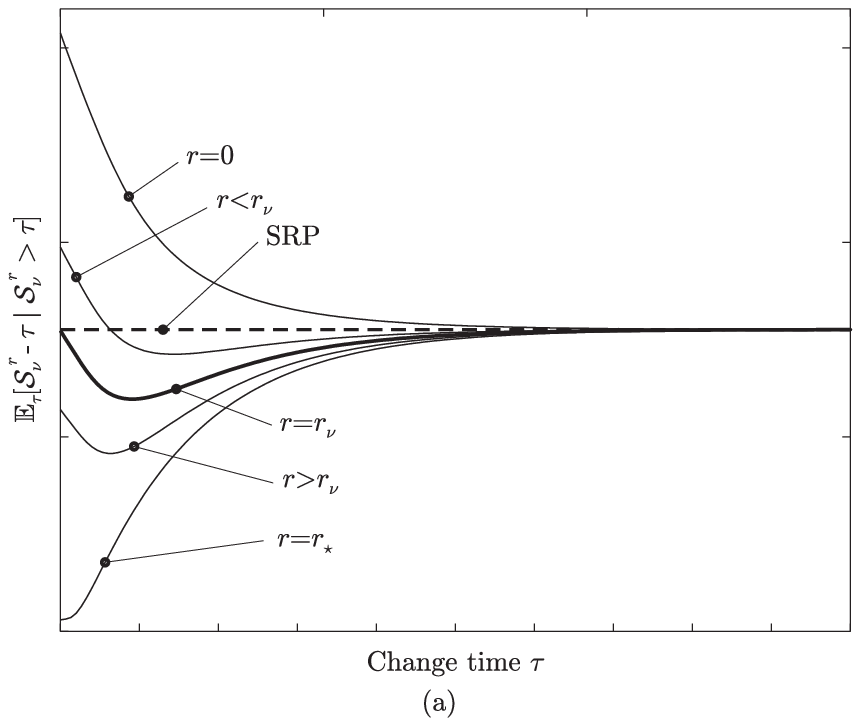}~\includegraphics[width=0.49\textwidth]{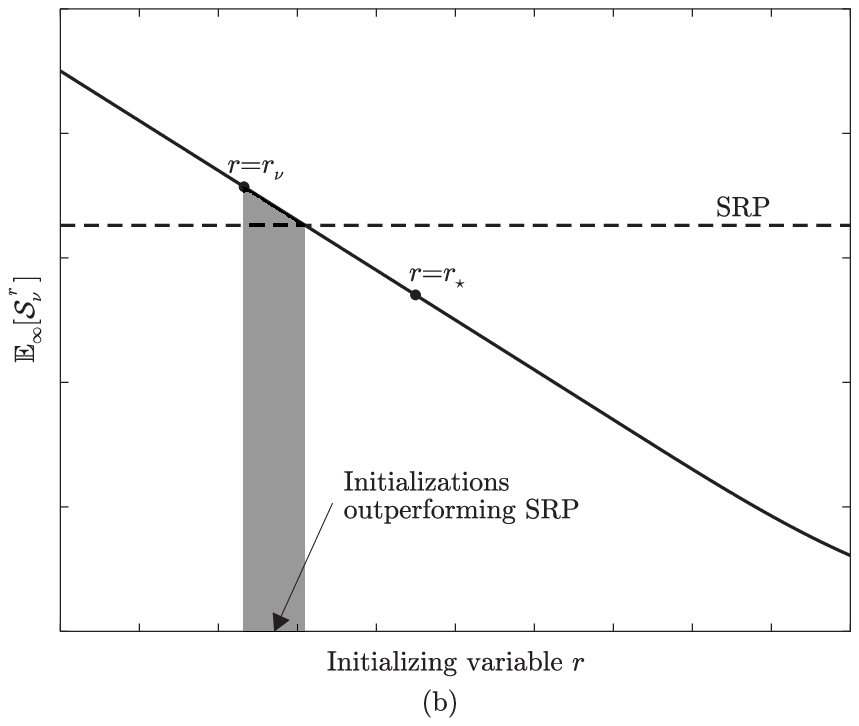}}
 \caption{\baselineskip0.4cm (a) Typical form of expected detection delay as a
 function of change-time $\tau$ for various initialization strategies and
 (b) ARL to false alarm as a function of the initializing parameter $r$. \label{f:Fig1}}
\end{figure}

It turns out that the proposed initialization strategy $r_\nu$ defined in \eqref{eq:opt_r} has the following property that can be used as a simpler, alternative definition. If we
fix $\nu$ and compute $\Exp_{\tau}[\cS_\nu^{r}-\tau|\cS_\nu^{r}>\tau]$,
$r_\nu$ is {\em the smallest} $r$ for which the supremum $\sup_{\tau\ge0}
\Exp_{\tau}[\cS_\nu^{r}-\tau|\cS_\nu^{r}>\tau]$ becomes equal to the steady state
value $\lim_{\tau\to\infty} \Exp_{\tau}[\cS_\nu^{r}-\tau|\cS_\nu^{r}>\tau]$.
Typical forms of $\Exp_{\tau}[\cS_\nu^{r}-\tau|\cS_\nu^{r}>\tau]$, as functions of
the change time $\tau$ and for different values of the initializing parameter $r$,
are depicted in Fig.\,\ref{f:Fig1}(a). As we can see, if $r<r_\nu$, then the
supremum of this function exceeds its steady state limit; whereas for $r\ge r_\nu$,
the supremum coincides with the steady state limit. We observe that
$\Exp_{\tau}[\cS_\nu^{r_\nu}-\tau|\cS_\nu^{r_\nu}>\tau]$ attains the steady state
value not only in the limit as $\tau\to\infty$, but also for some finite value of
$\tau$. In the same figure we can also see that the selection $r=0$, corresponding to the classical SR test, exhibits a decreasing behavior, with the worst detection delay appearing at $\tau=0$. On the other hand, the SRP test with threshold $\nu$ has a constant performance (dashed line) which coincides with the steady state value. Finally, we plot the expected detection delay for $r=r_\star$, where $r_\star$ is the smallest $r$ for which $\Exp_{\tau}[\cS_\nu^{r}-\tau|\cS_\nu^{r}>\tau]$ becomes an increasing
function of $\tau$.

Therefore, we conclude that, for any given threshold $\nu$, the
proposed initializing parameter $r_\nu$ can be alternatively defined as
\begin{equation}
r_\nu=\inf\left\{r\ge0:\Exp_{\tau}[\cS_\nu^r-\tau|\cS_\nu^r>\tau]\le\lim_{\tau\to\infty}
\Exp_{\tau}[\cS_\nu^r-\tau|\cS_\nu^r>\tau]\right\}.
\label{eq:rnu_alt}
\end{equation}
The advantage here is that this does not involve the computation of the lower bound
$\cL_{\rm P}(\cS_\nu^r)$.

In Fig.\,\ref{f:Fig1}(b) we plot the ARL to false alarm $\Exp_\infty[\cS_\nu^r]$ as
a function of the initializing parameter $r$. For the same threshold, the proposed SR-$r_\nu$ test exhibits a {\it
larger} ARL to false alarm than the one obtained by the SRP test. Since both tests
have the same worst conditional average detection delay (i.e., the same Pollak
measure), this suggests that initialization with $r=r_\nu$ is preferable to the SRP.
Note that all values of $r$ inside the
half-toned strip in Fig.\,\ref{f:Fig1}(b) correspond to tests that perform better
than the SRP procedure. Furthermore, since the corresponding values of $r$ are larger than
$r_\nu$, their worst detection delay is equal to the steady
state value and thus equal to the SRP performance.

As a last remark, recall
that $r_\star$ corresponds to the smallest $r$ for which the conditional average
detection delay becomes an increasing function of $\tau$. Thus this initialization strategy
exhibits a strong fast initial response, responding much faster to changes that take place in earlier than later
stages, complete opposite to the classical SR test that
starts from $r=0$ and prefers large change-times $\tau$.

\null\par

\setcounter{chapter}{3}
\setcounter{equation}{0} 
\noindent {\bf 3. Proposed Methodology}
\\
We derive here {\em exact}
integral equations, as well as relevant recursive formulas, for the performance
metrics of the SR-$r$ and SRP tests. The exact formulation is then followed by a
set of approximations leading to classical linear problems from Linear Algebra that
can be solved numerically, and thus provide answers to long standing performance
evaluation problems.

\vskip0.4cm\par
\noindent {\bf 3.1 Integral Equations for the Operating Characteristics of the SR-$\boldsymbol{r}$ Test}
\\
Fix $r,\nu$ with $r\in[0,\nu)$ and define $\phi_i(r)=\Exp_i[\cS_\nu^r]$, where
$i=0,\infty$. The function $\phi_\infty(r)$ is the ARL to false alarm and $\phi_0(r)$
is the average detection delay when the change takes place before surveillance
begins. Since the statistic $R_n^r$ obeys the recursion $R_n^r=(1+R_{n-1}^r)\ell_n$
(cf.~\eqref{eq:statistics2}), it is clear that $\{R_n\}_{n\ge0}$ is a homogeneous
Markov process and, therefore, one has
$$
\phi_i(r)=1+\Exp_i\left[\phi_i(R_1^r)\indicator{R_1^r<\nu}|R_0^r=r\right],
$$
where, hereafter, $\mathbbm{1}_{\mathcal{A}}$ stands for the indicator of a set
$\mathcal{A}$. Since
$$
\Pro_i[R_1^r\le x|R_0^r=r]=
\Pro_i\left[\ell_1\le\frac{x}{1+r}\right]=F_i\left(\frac{x}{1+r}\right),
$$
where $F_i(\cdot)$ is the cdf of the likelihood ratio $\ell_1$ under the $\Pro_i$
measure, by substituting this equality into the previous one we obtain the final
integral form for the functions of interest
\begin{align}\label{eq:ARL-ADD-eqn-int-form}
\phi_i(r) = 1+\int_0^\nu\phi_i(x) \frac{\partial}{\partial
x}F_i\left(\frac{x}{1+r}\right)\,dx,\quad i=0,\infty.
\end{align}

The next important performance metric is the conditional expected detection delay
\begin{equation}
\Exp_\tau[\cS_\nu^r-\tau|\cS_\nu^r>\tau] =
\frac{\Exp_\tau[(\cS_\nu^r-\tau)^+]}{\Pro_\infty[\cS_\nu^r>\tau]},\quad
\tau=0,1,2\ldots \, . \label{eq:add_sr}
\end{equation}
We consider the numerator and the denominator separately and
propose recursive formulas for their computation.
For $\tau\ge 1$, let
$
\delta_\tau(r)=\Exp_\tau[(\cS_\nu^r-\tau)^+]~~\text{and}~~\rho_\tau(r)
=\Pro_\infty[\cS_\nu^r>\tau]
$. Due to the
Markov nature of $\{R_n^r\}_{n\ge0}$, for $\tau\ge1$, we have
recursions for the sequences of functions:
\begin{align}
\begin{split}
\delta_\tau(r)&=\Exp_\infty[\delta_{\tau-1}(R_1^r)\indicator{R_1^r<\nu}|R_0^r=r]\\
&=\int_0^{\nu}\delta_{\tau-1}(x)\,\frac{\partial}{\partial
x}F_\infty\left(\frac{x}{1+r}\right)\,dx
\end{split}
\label{eq:recursion1}
\end{align}
\begin{align}
\begin{split}
\rho_\tau(r)&=\Exp_\infty[\rho_{\tau-1}(R_1^r)\indicator{R_1^r<\nu}|R_0^r=r]\\
&=\int_0^{\nu}\rho_{\tau-1}(x)\,\frac{\partial}{\partial x}F_\infty\left(\frac{x}{1+r}\right)\,dx,
\end{split}
\label{eq:recursion2}
\end{align}
where $\delta_0(r)=\phi_0(r)$ and $\rho_0(r)=1$. The
sequences of functions
$\{\delta_\tau(r)\},\{\rho_\tau(r)\}$ can be computed using these
recursive formulas; they involve a repetitive application of the same linear
transformation on an initial function and
$$
\Exp_\tau[\cS_\nu^r-\tau|\cS_\nu^r>\tau]=\frac{\delta_\tau(r)}{\rho_\tau(r)},~~
\cJ_{\rm P}(\cS_\nu^r)=\sup_{\tau\ge0}\frac{\delta_\tau(r)}{\rho_\tau(r)} .
$$

\noindent{\bf Remark.} It is of interest to evaluate the local false alarm probabilities
$\Pro_\infty[\cS_\nu^r \le k+m | \cS_\nu^r > k]$, $k=0,1,\dots,$ inside a fixed
``window'' of size $m$ ($m\ge 1$; for $m=1$ we obtain the instantaneous false alarm
probability). In particular, the supremum local false alarm probability $\sup_{k
\ge 0}\Pro_\infty[\cS_\nu^r \le k+m | \cS_\nu^r > k]$ can serve as an alternative
measure of the false alarm rate in place of the ARL to false alarm (see Tartakovsky (2005, 2009) for a more detailed discussion). Since
$$
\Pro_\infty[\cS_\nu^r \le k+m | \cS_\nu^r > k] = \frac{\Pro_\infty[k < \cS_\nu^r
\le k+m]}{\Pro_\infty[\cS_\nu^r
> k]} = 1- \frac{\Pro_\infty[\cS_\nu^r > k+m]}{\Pro_\infty[\cS_\nu^r> k]},
$$
we obtain that
$
\Pro_\infty[\cS_\nu^r \le k+m | \cS_\nu^r > k] = 1- \rho_{k+m}(r)/\rho_{k}(r),
$
where $\rho_0(r) = 1$ and $\rho_j(r)$, $j= 1, 2 \dots$ are given in~\eqref{eq:recursion2}.

\vskip0.4cm\par
\noindent {\bf 3.2 Integral Equations for the Operating Characteristics of the SRP Test}
\\
We now derive equations for the operating characteristics of the randomized SRP
scheme. First, the
quasi-stationary density $q(x)$ satisfies the integral equation
\begin{equation}\label{eq:QSD-eqn-int-form}
\lambda_{\max}\, q(x)=\int_0^\nu q(r)\frac{\partial}{\partial
x}F_\infty\left(\frac{x}{1+r}\right)dr
\end{equation}
(cf.\ Pollak (1985)), where $\lambda_{\max}$ is the leading eigenvalue of the linear integral operator induced by the kernel
\begin{align*}
K_\infty(x,r)=\frac{\partial}{\partial x}F_\infty\left(\frac{x}{1+r}\right),~~
x,r\in[0,\nu),
\end{align*}
and, consequently, $q(x)$ is the corresponding (left) eigenfunction. Since $q(x)$ is
a probability density with support $[0,\nu)$, it also satisfies the constraint
\begin{equation}
\int_0^\nu q(x)\,dx=1.
\label{eq:int_to_1}
\end{equation}
Equations~\eqref{eq:QSD-eqn-int-form} and~\eqref{eq:int_to_1} are sufficient to
{\em uniquely define} $\lambda_{\max}$ and $q(x)$, while  their
existence is guaranteed by Harris (1963), Theorem III.10.1. Furthermore,
integrating both sides of~\eqref{eq:QSD-eqn-int-form} with respect to $x$ over the
interval $[0,\nu)$, using~\eqref{eq:int_to_1} and the fact that $F_\infty(0)=0$,
we conclude that
$$
0\le\lambda_{\max}=\int_0^\nu
q(r)F_\infty\left(\frac{\nu}{1+r}\right)\,dr<
\int_0^\nu
q(r)\,dr=1.
$$
Note that $F_\infty(\nu/(1+r))\le1$, but there is an interval
for $r$ where this inequality is strict thus implying
the strict inequality in the previous relation. Indeed, by assuming that the
pdfs before and after the change are different, we have that $F_\infty(1)<1$. Using
the continuity of $F_\infty(x)$ (as a result of the assumption that $\ell_1$ is
continuous) we have that $F_\infty(x)<1$ for $x\le1$ and sufficiently close to 1.
Now note that, for sufficiently large $\nu$ and $r$ sufficiently close to $\nu$, we
assess that $\nu/(1+r)<1$ and that it is sufficiently close to 1, therefore
$F_\infty(\nu/(1+r))<1$. Thus
the leading eigenvalue $\lambda_{\max}$ is nonnegative and strictly bounded by 1, and the same is true
for $F_0(x/(1+r))$.

Assuming that $q(x)$ is available through a solution of~\eqref{eq:QSD-eqn-int-form}, we proceed with the computation of the performance of $\cS_\nu^q$. Since $\cS_\nu^q$ is an equalizer rule, $\cJ_{\rm P}(\cS_\nu^q)=\bar{\Exp}_0[\cS_\nu^q]$, while the ARL to false alarm becomes $\bar{\Exp}_\infty[\cS_\nu^q]$. In the case of the SRP test, averaging is with respect not only to the observation statistics, but also to the distribution of the initializing point. The expected values $\bar{\Exp}_\infty[\cS_\nu^q]$ and
$\bar{\Exp}_0[\cS_\nu^q]$ are easy to compute when the quasi-stationary pdf $q(x)$
and the two functions $\phi_i(r),~i=0,\infty,$ are available. Indeed,
\begin{align}
\begin{split}
\bar{\Exp}_i[\cS_\nu^q]&=\int_0^\nu\Exp_i[\cS_\nu^q|R_0^q=x] q(x)dx
=\int_0^\nu\Exp_i[\cS_\nu^x] q(x)dx\\
&=\int_0^\nu\phi_i(x)q(x)dx.
\label{eq:srp_perf}
\end{split}
\end{align}

\vskip0.4cm\par
\noindent {\bf 3.3 Lower Bound Computation}
\\
Using the definition of $\delta_\tau(r)$ in~\eqref{eq:recursion1} and the fact that the leading eigenvalue of the linear transformation that updates the sequence of functions $\{\delta_\tau(r)\}$ is $\lambda_{\max}$, we conclude that
$\delta_\tau(r)=O(\lambda_{\max}^\tau)$. Since $0<\lambda_{\max}<1$, it follows
that the series in the numerator of the lower bound $\sum_{\tau=0}^\infty
\Exp_\tau[(\cS_\nu^r-\tau)^+]=\sum_{\tau=0}^\infty\delta_\tau(r)$ is absolutely
summable. Consequently, it is easy to verify that
$\psi(r)=\sum_{\tau=0}^\infty\delta_\tau(r)$ is the solution of the
integral equation
\begin{equation}
\psi(r)=\phi_0(r)+\int_0^{\nu}\psi(r)\,\frac{\partial}{\partial
x}F_\infty\left(\frac{x}{1+r}\right)\,dx.
\label{eq:tildephi0}
\end{equation}
Using (\ref{eq:bound}) and the above notation, we obtain
\begin{equation}
\cL_{\rm P}(\cS_\nu^r)=\frac{r\phi_0(r)+\psi(r)}{r+\phi_\infty(r)},
\label{eq:lower_bound}
\end{equation}
where $\phi_i(r)$, $i=0,\infty$ are given by~\eqref{eq:ARL-ADD-eqn-int-form} and
$\psi(r)$ by~\eqref{eq:tildephi0}.

\vskip0.4cm\par
\noindent {\bf 3.4 Numerical Solutions}
\\
Observe first that~\eqref{eq:ARL-ADD-eqn-int-form} and~\eqref{eq:QSD-eqn-int-form} can be written in the form
\begin{align}\label{eq:fredholm-eqn-2nd-kind}
u(r)-\alpha\int_0^\nu K(x,r)u(x)\,dx=v(r),
\end{align}
where $u(x)$ is an unknown function, $\alpha\neq 0$ and $x,r\in[0,\nu]$. In
particular, \eqref{eq:ARL-ADD-eqn-int-form} can be obtained
from~\eqref{eq:fredholm-eqn-2nd-kind} by letting $u(x)=\phi_i(x)$,
$K(x,r)=\frac{\partial}{\partial x}F_i(x/(1+r))$, $i=0,\infty$, $v(r)= 1$, and
$\alpha=1$. Replacing integration over $x$ in~\eqref{eq:fredholm-eqn-2nd-kind} with integration over $r$, and $u(x)$ with
$u(r)=q(r)$ under the integral and choosing $v(r)= 0$,
$K(x,r)=\frac{\partial}{\partial x}F_\infty(x/(1+r))$, and
$\alpha=1/\lambda_{\max}$, yields~\eqref{eq:QSD-eqn-int-form}. We assume
sufficient smoothness of our functions so that the interval $[0,\nu)$ can be
extended to $[0,\nu]$.

An equation analogous to~\eqref{eq:fredholm-eqn-2nd-kind} occurs in a wide variety of
physical applications and is known as the Fredholm equation of the second kind; see, e.g., Petrovskii (1957)
and Kress (1989). The fundamental result concerning the existence and uniqueness of
solutions of such equations is that, given certain regularity conditions on the
kernel, these equations have unique solutions provided $1/\alpha$ is not a {\em
proper} number or an eigenvalue of the linear integral operator associated with the kernel $K(x,y)$. As we have seen,
$\alpha=1$ is not an eigenvalue of either of the operators induced by $\frac{\partial}{\partial x}F_i(x/(1+r))$, $i=0,\infty$.

Various numerical schemes for solving~\eqref{eq:fredholm-eqn-2nd-kind} are
developed in Kantorovich and Krylov (1958), Petrovskii (1957), and Atkinson and Han
(2001). Commonly, one replaces the
function $f(r)=\int_0^\nu K(x,r)u(x)dx$
in~\eqref{eq:fredholm-eqn-2nd-kind} by a vector
$\boldsymbol{f}=[f(r_0),f(r_1),\ldots,f(r_N)]^t$, where $0=r_0<r_1<\cdots<r_N=\nu$
constitutes a sampling of the interval $[0,\nu]$. A similar sampling is
applied to the function $u(x)$ producing the vector
$\boldsymbol{u}=[u(x_0),u(x_1),\ldots,u(x_N)]^t$. The integral is then evaluated
using some numerical integration technique, leading to a
(right) matrix-vector multiplication that replaces the integral,
\begin{equation}
f(r)=\int_0^\nu K(x,r)u(x)dx \Rightarrow
\tilde{\boldsymbol{f}}=\boldsymbol{K}\boldsymbol{u},
\label{eq:app_int_r}
\end{equation}
where $\boldsymbol{K}$ is a matrix that depends on the numerical integration method
and the sampling points $\{r_i\},\{x_i\}$ and
$\tilde{\boldsymbol{f}}=[\tilde{f}(r_0),\tilde{f}(r_1),\ldots,\tilde{f}(r_N)]^t$,
with $\tilde{f}(r)$ denoting the approximation to $f(r)$ as a result of evaluating the integral numerically. The {\it same}
matrix $\boldsymbol{K}$ used for the numerical evaluation of the integral
in~\eqref{eq:app_int_r} can also be used to evaluate the conjugate integral via
\begin{align}
f(x)=\int_0^\nu K(x,r) u(r)dr \Rightarrow \tilde{\boldsymbol{f}}^t
=\boldsymbol{u}^t\boldsymbol{K}.
\label{eq:app_int_l}
\end{align}

To find the matrix $\boldsymbol{K}$ we need to use numerical integration. We employ the simplest numerical technique to demonstrate the main idea; if one adopts more powerful
numerical integration methods, the results will be of higher accuracy.

Consider an integral of the form $\int_a^b z(x)F'(x)\,dx$, where $F'(x)$ is the
derivative of $F(x)$ and let $a=x_0<x_1<\cdots<x_N=b$ be a sampling of the
interval $[a,b]$. Then
\begin{equation}
\begin{split}
\int_a^b z(x)F'(x)\,dx\approx&~\frac{1}{2}\sum_{j=1}^N [F(x_j)-F(x_{j-1})] [z(x_j)+z(x_{j-1})]\\
=&~\frac{1}{2}[F(x_1)-F(x_0)]z(x_0)\\
&~+\sum_{j=1}^{N-1}\frac{1}{2}[F(x_{j+1})-F(x_{j-1})]z(x_j)\\
&~+\frac{1}{2}[F(x_N)-F(x_{N-1})]z(x_N).
\end{split}
\label{eq:app_int}
\end{equation}

For~\eqref{eq:ARL-ADD-eqn-int-form} and~\eqref{eq:QSD-eqn-int-form}), as well as the computations in~\eqref{eq:recursion1} and~\eqref{eq:recursion2} that involve the kernels $K_i(x,r)=\frac{\partial}{\partial
x}F_i(\frac{x}{1+r}),~i=0,\infty$, sample the interval $[0,\nu]$ canonically at
$x_j=r_j=jc,~j=0,\ldots,N$ with $c=\nu/N$, and apply~\eqref{eq:app_int_r},~\eqref{eq:app_int_l} and~\eqref{eq:app_int} to obtain
\begin{align}
\int_0^\nu \frac{\partial}{\partial
x}F_i\left(\frac{x}{1+r}\right)u(x)\,dx&\Rightarrow
\boldsymbol{M}_i\boldsymbol{u},\label{1st_app}\\
\int_0^\nu \frac{\partial}{\partial
x}F_i\left(\frac{x}{1+r}\right)u(r)\,dr&\Rightarrow
\boldsymbol{u}^t\boldsymbol{N}_i.\label{2nd_app}
\end{align}
The matrices $\boldsymbol{M}_i,\boldsymbol{N}_i$ are of size $(N+1)\times(N+1)$ with elements
\begin{align}
(\boldsymbol{M}_i)_{k,m}&=\left\{
\begin{array}{cl}
0.5F_i\left(\frac{c}{1+mc}\right)& \text{for $k=0$},\\
0.5F_i\left(\frac{(k+1)c}{1+mc}\right)-0.5F_i\left(\frac{(k-1)c}{1+mc}\right) & \text{for $N> k\ge 1$},\\
0.5F_i\left(\frac{\nu}{1+mc}\right)-0.5F_i\left(\frac{(N-1)c}{1+mc}\right) &
\text{for $k=N$},
\end{array}
\right.
\label{eq:matrix_M}
\end{align}
where in the first line of~\eqref{eq:matrix_M} we used the fact that
$F_i(0)=0$; and
\begin{align}
(\boldsymbol{N}_i)_{m,k}&=\left\{
\begin{array}{cl}
0.5F_i\left(mc\right)-0.5F_i\left(\frac{mc}{1+c}\right)& \text{for $k=0$},\\
0.5F_i\left(\frac{mc}{1+(k+1)c}\right)-0.5F_i\left(\frac{mc}{1+(k-1)c}\right) & \text{for $N> k\ge 1$},\\
0.5F_i\left(\frac{mc}{1+\nu}\right)-0.5F_i\left(\frac{mc}{1+(N-1)c}\right) &
\text{for $k=N$}.
\end{array}
\right.
\label{eq:matrix_N}
\end{align}

Using~\eqref{1st_app}, \eqref{eq:ARL-ADD-eqn-int-form} and~\eqref{eq:tildephi0} are
reduced to
\begin{align}
\begin{split}
\tilde{\bphi}_i&=J+\boldsymbol{M}_i\tilde{\bphi}_i,~i=0,\infty,\\
\tilde{\boldsymbol{\psi}}&=\tilde{\bphi}_0+\boldsymbol{M}_\infty\tilde{\boldsymbol{\psi}},
\end{split}
\label{eq:sample_phi}
\end{align}
where $\tilde{\bphi}_i=[\tilde{\phi}_i(0),\tilde{\phi}_i(c), \cdots,
\tilde{\phi}_i(\nu)]^t$,
$\tilde{\boldsymbol{\psi}}=[\tilde{\psi}(0),\tilde{\psi}(c), \cdots,
\tilde{\psi}(\nu)]^t$, with $\tilde{\phi}_i(x)$, $\tilde{\psi}(x)$ denoting the
approximation to $\phi_i(x)$ and $\psi(x)$, respectively, and $J=[1\cdots1]^t$.
Solving the linear system of equations in~\eqref{eq:sample_phi} yields the required
approximation for the functions $\phi_i(r)$ and $\psi(r)$.

The recursions in~\eqref{eq:recursion1} and~\eqref{eq:recursion2}, using
again~\eqref{1st_app}, can be approximated by
\begin{align}
\tilde{\bdelta}_{\tau}& =\boldsymbol{M}_\infty\tilde{\bdelta}_{\tau-1},~\tilde{\bdelta}_0=\tilde{\bphi}_0,\\
\tilde{\brho}_{\tau}&=\boldsymbol{M}_\infty\tilde{\brho}_{\tau-1},~\tilde{\brho}_0=J,
\end{align}
where $\tilde{\bdelta}_{\tau}=[\tilde{\delta}_{\tau}(0),\tilde{\delta}_{\tau}(c),
\tilde{\delta}_{\tau}(2c), \cdots,\tilde{\delta}_{\tau}(\nu)]^t$, $c=\nu/N$, and
$\tilde{\delta}_{\tau}(x)$ denotes the approximation to $\delta_{\tau}(x)$. A
similar definition applies to $\tilde{\brho}_{\tau}$.

We now turn to~\eqref{eq:QSD-eqn-int-form}. Since this involves
conjugate integration, we need to apply the approximation given in~\eqref{2nd_app},
which leads to
\begin{equation}
\tilde{\lambda}_{\max}\tilde{\boldsymbol{q}}^t
=\tilde{\boldsymbol{q}}^t\boldsymbol{N}_\infty.
\label{eq:eig-eig}
\end{equation}
This suggests that $(\tilde{\lambda}_{\max},\tilde{\boldsymbol{q}})$ is a left
eigenvalue-eigenvector pair for the matrix $\boldsymbol{N}_\infty$ with
$\tilde{\lambda}_{\max}$ being the leading eigenvalue of $\boldsymbol{N}_\infty$. Of course $\tilde{\boldsymbol{q}}$ is not unique unless we use the
constraint~\eqref{eq:int_to_1}. Applying~\eqref{eq:app_int}, this constraint is
transformed to
\begin{align}
c[0.5,~1,\cdots,1,~0.5]\tilde{\boldsymbol{q}}=1,
\end{align}
where $c=\nu/N$. Since the matrix
$\boldsymbol{N}_\infty$ has positive elements, see
\eqref{eq:matrix_N}, its leading eigenvalue $\tilde{\lambda}_{\max}$
and the corresponding left and right eigenvectors (consequently also
$\tilde{\boldsymbol{q}}$) are necessarily nonnegative (see, e.g., Horn and Johnson
(1990)). Following similar arguments as in Subsection\,3.2 we can
show that $0\le\tilde{\lambda}_{\max}<1$.

For computing the performance of the SRP procedure from~\eqref{eq:srp_perf} we have
\begin{equation}
\bar{\Exp}_i[\cS_\nu^q]\approx c[0.5,~1,\cdots,1,~0.5](\tilde{\boldsymbol{\phi}}_i\circ\tilde{\boldsymbol{q}}), \label{eq:last}
\end{equation}
where, if $\boldsymbol{x},\boldsymbol{y}$ are vectors of the same length,
$\boldsymbol{x}\circ\boldsymbol{y}$ denotes the vector that
results from the  element-by-element multiplication.

Now (\ref{eq:sample_phi}) -- (\ref{eq:last}) can be used to obtain numerical
solutions for our performance evaluation problem. When $\nu$ is large and/or
sampling is fine, solving~\eqref{eq:sample_phi} and~\eqref{eq:eig-eig} can be
particularly memory and time demanding. For such large-sized problems it is
preferable to apply iterative solution techniques that avoid storage of the
matrices $\boldsymbol{M}_i,\boldsymbol{N}_i$ and, for Eq.\,\eqref{eq:sample_phi}, 
to use simple pre-conditioning techniques to speed up convergence (see, e.g., Quarteroni, Sacco, and Saleri (2000)).

Approximation accuracy is, of course, directly related to $N$. If $N$ is sufficiently large, the numerical
solution is close to exact, see Kantorovich and Krylov (1958) or Atkinson and
Han (2001). More details on computation of efficient upper bounds for the numerical errors are reported in Polunchenko (2009).

In the next section, we apply these ideas in specific examples in order to compare
the relative performance of different initialization strategies.

\null\par
\setcounter{chapter}{4}
\setcounter{equation}{0} 
\setcounter{figure}{0}
\noindent {\bf 4. Numerical Examples}
\\
Apart from the initialization strategies introduced in Subsection\,2.1, we also examine the case of SR-$\mu$ with $r=\mu=\int_0^\nu x q(x) dx$, where we initialize the test with the mean of the quasi-stationary distribution. Recall that SR-$r_\star$ corresponds to the smallest value of $r$ for which $\Exp_\tau[\cS_\nu^r-\tau|\cS_\nu^r>\tau]$ becomes increasing with respect to $\tau$ (for fast initial response) and SR-$r_\nu$ is the initialization obtained by solving~\eqref{eq:opt_r} or~\eqref{eq:rnu_alt}. We did a comparative study of the change detection procedures for two models -- Gaussian and Exponential.

\vskip0.4cm\par
\noindent {\bf 4.1 Gaussian Example}
\\
Consider a Gaussian example of detecting a change in the mean value where
observations are i.i.d. $\gaussdf(0,1)$ pre-change and i.i.d. $\gaussdf(\theta,1)$
post-change:
\begin{align*}
f_\infty(x)=\frac{1}{\sqrt{2\pi}}\exp\set{-\frac{x^2}{2}}\quad\text{and}\quad
f_0(x)=\frac{1}{\sqrt{2\pi}}\exp\set{\frac{-(x-\theta)^2}{2}}, \quad \theta \neq 0.
\end{align*}
We performed extensive numerical computations for various parameter values, but
present only sample results for $\theta=0.1$, corresponding to a relatively small
change that is not easily detectable. We considered $\Exp_\infty[\cS_\nu^r]=10^3,
10^4$, as moderate and low false alarm rates, corresponding to detection
threshold $\nu$ in the range $1000\pm 10\%$ and $10000\pm 10\%$, respectively. 
The integration interval $[0,\nu]$ was sampled at $N=10^4$ and
$N=10^5$ equidistant points. We believe that such sampling is sufficiently
fine since the results of Monte Carlo
experiments for the conventional SR procedure (with $10^6$ replications) matched
our numerical results within $0.5\%$.

It is important to have a fairly accurate initial guess in order to
obtain a pilot estimate of $\Exp_\infty[\cS_\nu^r]$ in searching for appropriate
threshold values in a relatively narrow interval. To this end, the
approximation $\Exp_\infty[\cS_\nu^r]=\nu/w-r$ is used, where the constant
$w\in(0,1)$ (related to the ``overshoot'') is the subject of renewal theory and can
be computed numerically. This approximation can be obtained by noticing that $R_n^r
-n-r$ is a $\Pro_\infty$-martingale with zero expectation. Consequently, by the
optional sampling theorem, we have $\Exp_\infty[R_{\cS_\nu^r}^r-\cS_\nu^r -r]=0$.
Hence $\Exp_\infty[\cS_\nu^r]=\Exp_\infty[R_{\cS_\nu^r}^r] -r$ and, since
$R_{\cS_\nu^r}^r$ is the first excess over $\nu$, renewal theory can be applied to the ``overshoot'' $\log(R_{\cS_\nu^r}^r)-\log\nu$. This approximation was first derived for $r=0$ in Pollak (1987), and its generalization for any $r\in (0, \nu)$ is straightforward. For the SRP procedure the value of $r$ should be replaced by $\mu=\Exp_\infty[R_0^q]$, the mean of the quasi-stationary distribution.

\begin{figure}
 \centerline{\includegraphics[width=0.49\textwidth]{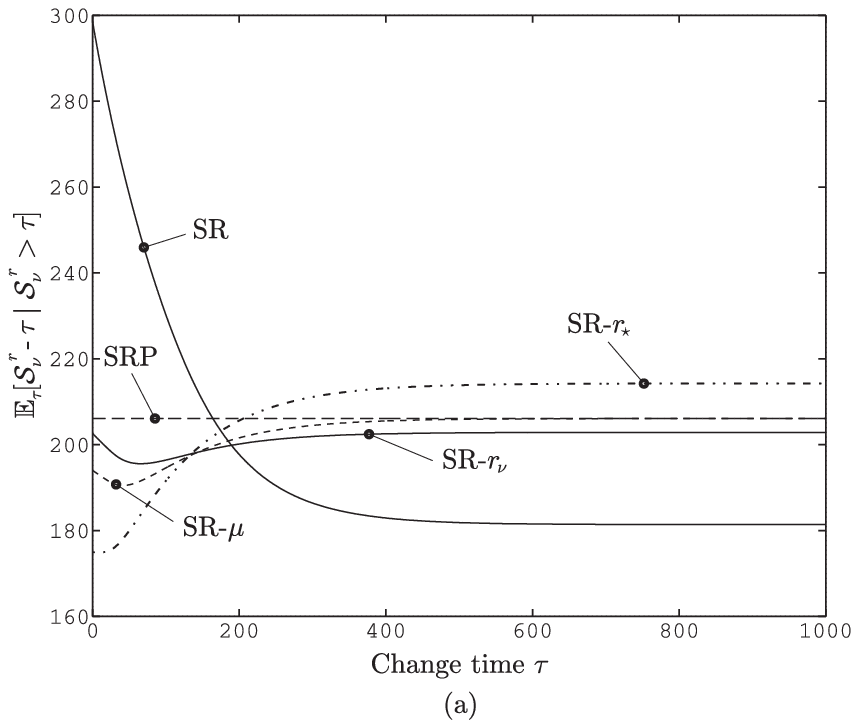}~\includegraphics[width=0.49\textwidth]{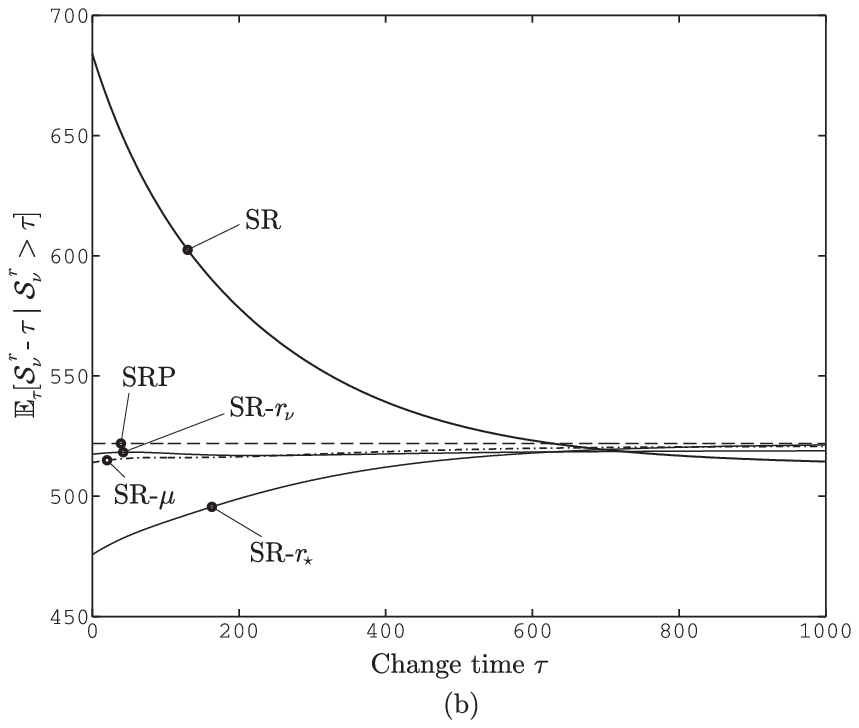}}
 \caption{\baselineskip0.4cm Average detection delay $\Exp_\tau[\cS_\nu^r-\tau|\cS_\nu^r>\tau]$
 for the different initialization strategies as a function of $\tau$ for $\theta=0.1$ and for ARL to false alarm:
(a)~$\gamma=10^3$, (b)~$\gamma=10^4$. \label{fig:Fig2}} 
\end{figure}

Fig.\,\ref{fig:Fig2}(a) shows the family of curves
$\Exp_\tau[\cS_\nu^r-\tau|\cS_\nu^r>\tau]$ versus $\tau$ for all initialization procedures in question when $\theta=0.1$ and ARL to false alarm $\gamma=10^3$. Fig.\,\ref{fig:Fig2}(b) depicts the same curves for $\gamma=10^4$. In order to have a more precise idea of the relative performance difference of the competing schemes, in Table\,\ref{tab:values} we list the numerical values obtained by our computational method for characteristic values $\tau$ and the ARL to false alarm of $10^3$. Table\,\ref{tab:thresholds} depicts the thresholds and the corresponding values of the initializing parameter $r$ that assure the desired values of the ARL to false alarm for each initialization strategy.

\begin{table}[!htb]
    \centering
    \caption{ Average detection delay $\Exp_\tau[\cS_\nu^r-\tau|\cS_\nu^r>\tau]$
 versus change point $\tau$ for the ARL to false alarm $\gamma=10^3$ and $\theta=0.1$. \label{tab:values}}
 \smallskip
    \begin{tabular}{|l||c|c|c|c|c|c|c|c|}
    \hline Test$\backslash\tau$
        &0&50&100&200&400&600&800&1000\\
        \hline
    \hline SR
        &298.5 &258.3 &230.2 &197.7 &182.9 &181.5 &181.4 &181.4\\
    \hline SR-$r_\nu$
        &202.8 &195.9 &196.4 &200.1 &202.5 &202.8 &202.8 &202.8\\
    \hline SR-$r_\star$
        &174.9 &179.9 &191.6 &205.6 &213.1 &214.1 &214.2 &214.3\\
    \hline SR-$\mu$
        &194.0 &190.7 &194.6 &201.6 &205.6 &206.0 &206.1 &206.1\\
    \hline SRP
        &\multicolumn{8}{c|}{206.1}\\
        \hline
    \end{tabular}
 \end{table}

\begin{table}[!htb]
    \centering
    \caption{Detection thresholds and initializing parameters resulting in the ARL to
 false alarm $\gamma=10^3,~10^4$ for $\theta=0.1$. \label{tab:thresholds}}
\smallskip
    \begin{tabular}{|l||r|r||r|r|}
        \hline
    \multicolumn{1}{|c||}{$\gamma$}&\multicolumn{2}{c||}{$10^3$}&\multicolumn{2}{c|}{$10^4$}\\
        \hline
    \multicolumn{1}{|c||}{Test}&\multicolumn{1}{c}{$\nu$}&\multicolumn{1}{|c||}{$r$}&\multicolumn{1}{c|}{$\nu$}&\multicolumn{1}{|c|}{$r$}\\
        \hline
    \hline SR  &944.0 &0&9435.0 &0\\
    \hline SR-$r_\nu$ &1142.0 &210.8&9775.0 &361.2\\
    \hline SR-$r_\star$ &1258.0&333.2 &9792.0&380.4\\
    \hline SR-$\mu$ &1174.0 &244.4 &9945.0&540.9\\
    \hline SRP &1174.0 & random&9945.0& random\\
        \hline
    \end{tabular}
 \end{table}

As we can see, the SRP procedure maintains constant average detection delay as
expected. The SR-$r_\star$ test has the fastest initial response (for immediate and
early changes), but the worst minimax behavior. The SR-$r_\nu$ procedure is uniformly better than the SRP test. Even though the difference is not dramatic it is nonetheless
visible here. It is interesting to note that the SR-$\mu$ detection procedure has an
intermediate performance between SR-$r_\nu$ and SR-$r_\star$, namely sufficiently
fast initial response and the same minimax performance as the SRP test attained at steady state.

Regarding the conventional SR test (with $r=0$) note that it outperforms all
competing schemes including SRP for sufficiently large change-time $\tau$. This is
expected since, as can be seen from Fig.\,\ref{f:Fig1}(b), when all tests have the
same threshold the SR test has the largest ARL to false alarm and the same
steady-state value for the expected detection delay. In the other tests, in order
to attain the same ARL to false alarm value as SR, thresholds should be increased.
This will result in an increase in the expected detection delay and, in particular,
the corresponding steady state value. Consequently, the expected delay of SR, due
to its monotone behavior, attains smaller values than the other tests for
sufficiently large change-time $\tau$.

\begin{figure}
\centerline{\includegraphics[width=0.49\textwidth]{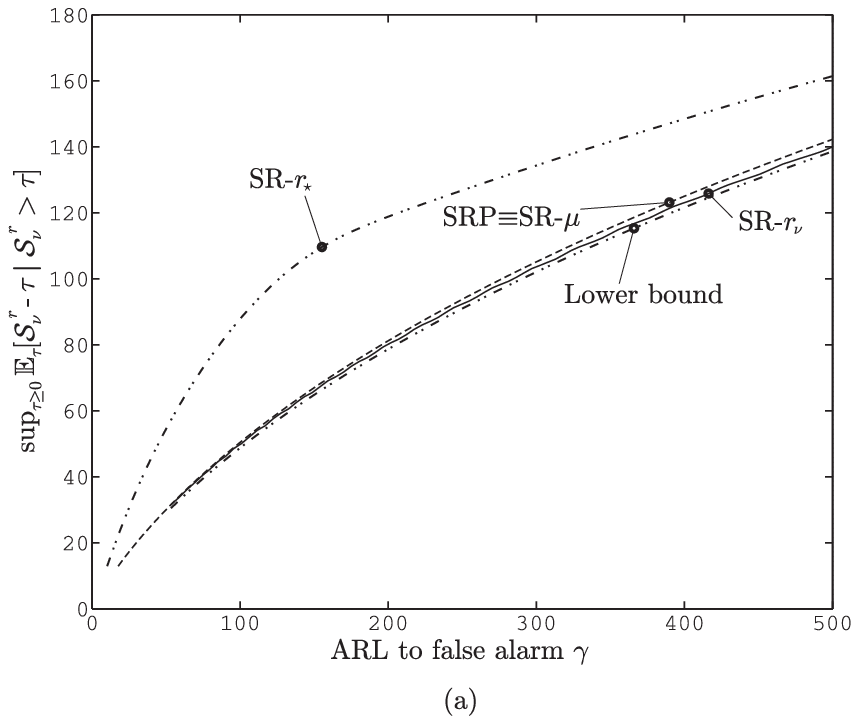}~\includegraphics[width=0.49\textwidth]{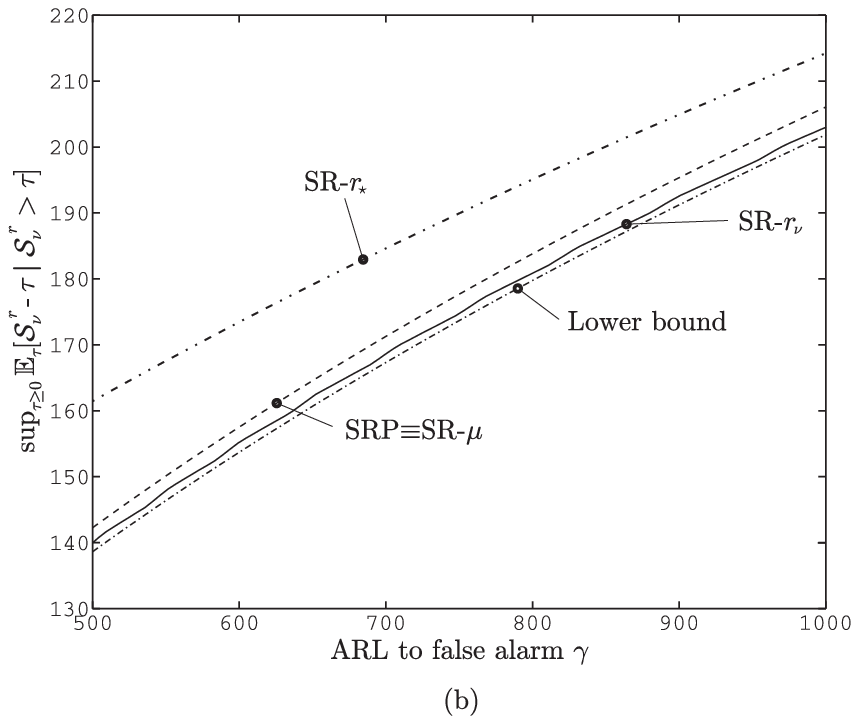}}
\caption{\baselineskip0.4cm Worst average detection delay of competing
initialization strategies and corresponding lower bound as a function of the ARL to
false alarm $\gamma$ for $\theta=0.1$, Gaussian case.\label{fig:Fig3}} 
\end{figure}

Fig.\,\ref{fig:Fig3} (a) and (b) depict the supremum average detection delay
$\cJ_{\rm P}(\cS_\nu^r)$ as a function of the ARL to false alarm
$\Exp_\infty[\cS_\nu^r]$ for the initialization strategies of interest, along with
the lower bound $\cL_{\rm P}(\cS_\nu^{r_\nu})$. We can see that the SR-$r_\nu$ test
uniformly outperforms all its rivals. We also observe that initializing the SR test
deterministically with the mean of the quasi-stationary distribution results in a performance that is indistinguishable from that of SRP.
Finally, we can see that the SR-$r_\star$ procedure is inferior to SRP, but we
recall that this version of the SR test has the best performance in terms of fast
initial response. Summarizing, the best minimax performance is delivered by the
SR-$r_\nu$ test. This performance is also very close to the lower bound $\cL_{\rm
P}(\cS_\nu^{r_\nu})$, suggesting that the unknown optimal test can offer only
insignificant improvement over SR-$r_\nu$.

\vskip0.4cm\par
\noindent {\bf 4.2 Exponential Example}
\\
Consider now the case where observations are independent, originally having an
Exponential(1) distribution, changing at an unknown time $\tau$ to
Exponential$(\theta)$,
\begin{equation} \label{Exppmodel}
f_\infty(x)= e^{-x}\indicator{x\ge 0}, ~~ f_0(x) = \theta^{-1}
e^{-\frac{x}{\theta}}\indicator{x\ge 0}, \quad \theta >0.
\end{equation}

\begin{figure}
  \centerline{\includegraphics[width=0.49\textwidth]{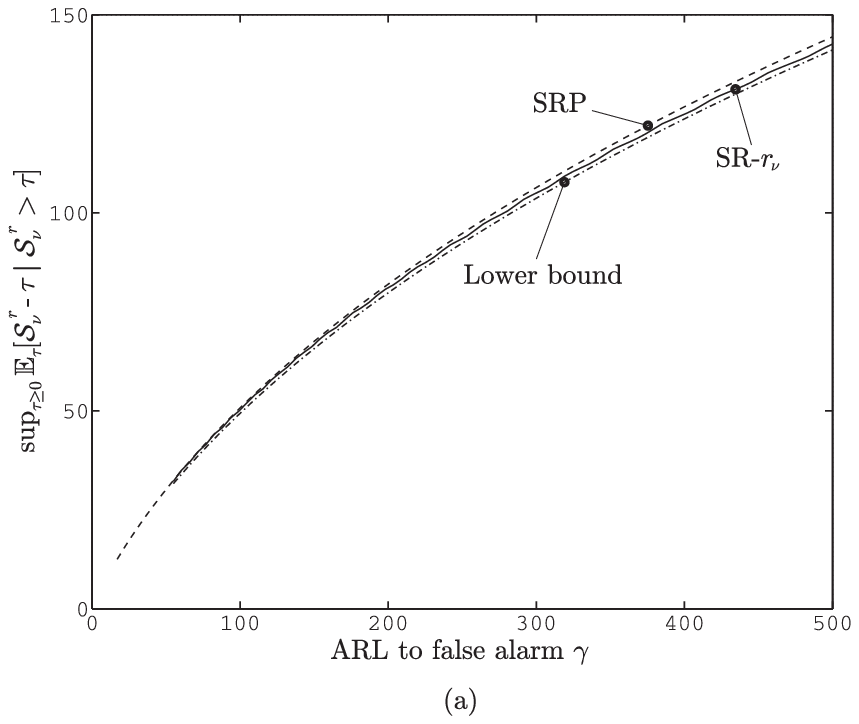}~\includegraphics[width=0.49\textwidth]{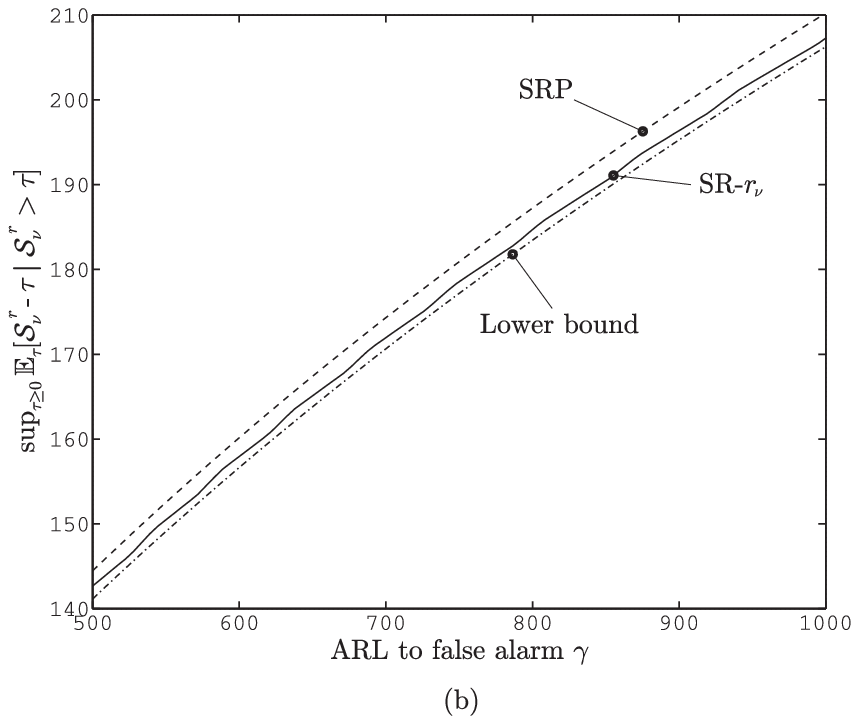}}
\caption{\baselineskip0.4cm Worst average detection delay of SRP and SR-$r_\nu$
and corresponding lower bound as a function of the ARL to false alarm $\gamma$ for
$\theta=1.1$, exponential case.\label{fig:Fig4}} 
\end{figure}
Fig.\,\ref{fig:Fig4} (a) and (b) depict the supremum average detection delay
$\cJ_{\rm P}(\cS_\nu^r)$ versus the ARL to false alarm for the SRP and SR-$r_\nu$
detection procedures, along with the lower bound $\cL_{\rm P}(\cS_\nu^{r_\nu})$ when
$\theta=1.1$. As in the Gaussian example, the operating characteristic of the
SR-$r_\nu$ procedure is uniformly better than that of the SRP procedure, with the
worst average detection delay $\cJ_{\rm P}(\cS_\nu^{r_\nu})$ being very close to
the lower bound of the minimax risk. Again we observe that there is very little
margin for improvement over the proposed detection procedure SR-$r_\nu$.

\vskip0.4cm\par
\noindent {\bf 4.3 Extensions for CUSUM and EWMA}
\\
Extending our results to cover the case of CUSUM and EWMA presents no special
difficulty. Consider first the CUSUM procedure defined by the CUSUM
statistic $V_n^r$ and the corresponding stopping time as
\[
V_n^r=\max\{1,V_{n-1}^r\}\ell_n,~~V_0^r=r; \quad
\cT_\nu^r=\inf\{n\ge1:V_n^r\ge\nu\},
\]
where $0\le r<\nu$. Note that the classical CUSUM is initialized with $r=1$ (cf.,
e.g., Moustakides (1986)). Here we consider the more general case suggested by Lucas
and Crosier (1982), which leads to fast initial response. When we compare~\eqref{eq:statistics2} and~\eqref{eq:st2} with the previous two formulas, we find a difference only in the term
$(1+R_{n-1}^r)$, which is now replaced by $\max\{1,V_{n-1}^r\}$. The same
difference is encountered in the integral equations that specify the two functions
$\phi_i(r)=\Exp_i[\cT_\nu^r],~i=0,\infty$, as well as the recursions that compute
the numerator and denominator of $\Exp_\tau[\cT_\nu^r-\tau|\cT_\nu^r>\tau]$.
Specifically,
\begin{align*}
\phi_i(r)&=1+\int_0^\nu\phi_i(x)\frac{\partial}{\partial
x}F_i\left(\frac{x}{\max\{1,r\}}\right)\,dx,~i=0,\infty,\\
\delta_\tau(r)&=\int_0^{\nu}\delta_{\tau-1}(x)\,\frac{\partial}{\partial
x}F_\infty\left(\frac{x}{\max\{1,r\}}\right)\,dx,~~\delta_0(r)=\phi_0(r),\\
\rho_\tau(r)&=\int_0^{\nu}\rho_{\tau-1}(x)\,\frac{\partial}{\partial
x}F_\infty\left(\frac{x}{\max\{1,r\}}\right)\,dx, ~~\rho_0(r)=1.
\end{align*}
A randomized version with $V_0$ following the
quasi-stationary distribution $\Pro[V_0 \le x]=\lim_{n\to\infty} \Pro[V_n^1 \le x |
\cT_\nu^1 >n]$ is also possible to define, with pdf satisfying
$$
\lambda_{\max}\, q(x)=\int_0^\nu q(r)\frac{\partial}{\partial
x}F_\infty\left(\frac{x}{\max\{1,r\}}\right)dr.
$$
In fact randomization is not necessary since the conventional CUSUM test with $r=1$ is exactly optimal in the sense of Lorden, \eqref{eq:lorden}, and the randomized CUSUM is always inferior to the SRP test in the sense of Pollak, \eqref{eq:pollak}.

Approximations that produce numerical solutions to these equations can be found in the same way.
Note that Dragalin (1994) used a slightly different, but very precise, numerical
technique for the computation of the ARL to false alarm $\Exp_\infty[\cT_\nu^1]$ and the average detection delay $\Exp_0[\cT_\nu^1]$ of the standard CUSUM for the Gaussian distribution. Comparison of results obtained by our numerical technique with those obtained by Dragalin shows that the two approximations are close, suggesting that our simple numerical method is of sufficiently high precision.

Finally, in order to make a similar extension for EWMA, here
\begin{align*}
D_n^r=\left(D_{n-1}^r\right)^\alpha \ell_n,~~D_0^r=r;\quad
\cN_{\nu_1,\nu_2}^r=\inf\{n\ge1:D_n^r\not\in(\nu_1,\nu_2)\},
\end{align*}
where $D_n^r$ is the EWMA statistic, $\cN_{\nu_1,\nu_2}^r$ is the corresponding
(double sided) stopping time, $0<\alpha<1$ is a forgetting factor, and
$0<\nu_1<1<\nu_2$ are the thresholds (the case $\nu_1=0$
corresponds to the one-sided EWMA procedure). Note that the EWMA statistic is
usually written in a form involving the log-likelihood ratio $\log(\ell_n)$. This
conventional form can be recovered by simply taking the logarithms, but we prefer the exponential version
above, since it allows for the derivation of integral equations for a variety of performance metrics.
Observing that the difference with the SR case is that the term
$(1+R_{n-1}^r)$ is replaced by $(D_{n-1}^r)^\alpha$, yields the
equations that define the operating characteristics
\begin{align*}
\phi_i(r)&=1+\int_{\nu_1}^{\nu_2}\phi_i(x)\frac{\partial}{\partial
x}F_i\left(\frac{x}{r^\alpha}\right)\,dx,~i=0,\infty,\\
\delta_\tau(r)&=\int_{\nu_1}^{\nu_2}\delta_{\tau-1}(x)\,\frac{\partial}{\partial
x}F_\infty\left(\frac{x}{r^\alpha}\right)\,dx,~~\delta_0(r)=\phi_0(r),\\
\rho_\tau(r)&=\int_{\nu_1}^{\nu_2}\rho_{\tau-1}(x)\,\frac{\partial}{\partial
x}F_\infty\left(\frac{x}{r^\alpha}\right)\,dx,
~~\rho_0(r)=1,\\
\lambda_{\max}\, q(x)&=\int_{\nu_1}^{\nu_2} q(r)\frac{\partial}{\partial
x}F_\infty\left(\frac{x}{r^\alpha}\right)dr,
\end{align*}
where the last expression corresponds to the pdf of the quasi-stationary
distribution defined as $\Pro[D_0 \le x]=\lim_{n\to\infty} \Pro_\infty [D_n^r \le
x|\cN_{\nu_1,\nu_2}^r>n]$ when a randomized EWMA is being constructed. 
To our knowledge, no randomized EWMA scheme has been previously considered.

Producing numerical approximations is again straightforward. Note that Robinson and Ho (1978)
proposed a different approach to obtaining numerical approximations for the performance of a somewhat different EWMA procedure. We believe that our approach is advantageous because it allows not only for the evaluation of the ARL to false alarm $\Exp_\infty [\cN_{\nu_1,\nu_2}^r]$ and the ARL to detection $\Exp_0 [\cN_{\nu_1,\nu_2}^r]$, but also for the optimization of the initializing parameter $r$. Furthermore, our method can be used to find the change time $\tau$ that produces the worst conditional expected detection delay. As opposed to the standard SR and CUSUM, in this test it is expected that this worst performance appears at a time $\tau>0$.

\null\par
\setcounter{chapter}{5}
\setcounter{equation}{0} 
\noindent {\bf 5. Conclusion}
\\
For the problem of quickest change detection with known pre- and post-change
distributions, we proposed a simple modification of the SR procedure, called the
SR-$r$ test, that starts from a deterministic (fixed) point $r \ge 0$. This
procedure represents a family of sequential tests, as the initializing point $r$
can take any value in the interval $[0,\nu)$, with $\nu$ denoting the threshold.
Our main contribution is the development of integral equations for
the major performance metrics of the SR-$r$ test and the corresponding numerical techniques for
solving these equations. Additionally, we give a method
for the numerical computation of the quasi-stationary distribution of the SR
statistic. This allows the practical implementation of
the randomized SRP procedure, introduced by Pollak (1985), that is known to enjoy
strong asymptotic optimality properties. Using our numerical methodology we 
can compute the operating characteristics of the conventional SR procedure
(that starts from 0) and of several interesting SR-$r$ variants corresponding to
specific choices of the initializing parameter $r$, and can compare their performance
against the SRP test.

The numerical results, obtained for the Gaussian and Exponential examples, indicate
that a specially designed SR-$r$ test can uniformly (for all points of change)
outperform the SRP procedure. Even though the difference is not dramatic, this
observation constitutes strong evidence against the exact optimality property of
the SRP procedure. As Polunchenko and Tartakovsky (2010) prove, using the integral equations presented in Subsections 3.1 and 3.2, the SRP procedure is indeed not strictly optimal, while the SR-$r$ procedure may be optimal in certain examples. Our numerical analysis also shows that by slightly sacrificing performance in the worst case detection delay sense, it is possible to design SR-$r$ tests that exhibit fast initial response (i.e., guarantee faster detection of changes that occur at early stages).

\null\par
\noindent {\large\bf Acknowledgment}
\\
This work was supported in part by the U.S.\ Army Research Office MURI grant
W911NF-06-1-0094 and by the U.S.\ National Science Foundation grant CCF-0830419 at
the University of Southern California.

\null\par
\noindent{\large\bf References}
\begin{description}
\item
Atkinson,~K.,~and~Han,~W. (2001). {\it Theoretical Numerical Analysis: A
Functional Analysis Framework}. Springer, New York.

\item
Baron, M. (2002). Bayes and asymptotically pointwise optimal stopping rules
for the detection of influenza epidemics. In {\it Case Studies in Bayesian
Statistics} {\bf 6} (C. Gatsonis, R. E. Kass, A. Carriquiry, A. Gelman, D. Higdon,
D. K. Pauler and I. Verdinelli eds.) 153-163. Springer, New York.

\item
Basseville,~M.,~and~Nikiforov,~I.V. (1993). {\it Detection of Abrupt Changes:
Theory and Application}. Prentice Hall, Englewood Cliffs, New Jersey.

\item
Beibel,~M. (1996). A note on Ritov's Bayes approach to the minimax property
of the CUSUM procedure. {\it Ann.\ Statist.} {\bf 24}, 1804-1812.

\item
Dragalin, V.V. (1994). Optimal CUSUM envelope for monitoring the mean of
normal distribution. {\it Economic Quality Control} {\bf 9}, 185-202.

\item
Feinberg,~E.A.,~and~Shiryaev,~A.N. (2006). Quickest detection of drift
change for Brownian motion in generalized Bayesian and minimax settings. {\it
Statistics \& Decisions} {\bf 24}, 445-470.

\item
Ferguson, T.S. (1967). {\it Mathematical Statistics: A Decision Theoretic
Approach}. Academic Press, New York.

\item
Galstyan, A., Mitra, S., and Cohen, P.R. (2007). Probabilistic plan tracking
and detection for intelligence analysis. {\it Joint Statistical Meetings (JSM)}, Salt Lake City.

\item
Horn,~R.A.,~and~Johnson,~C.R. (1990). {\it Matrix Analysis}. Cambridge
University Press.

\item
Harris, T.E. (1963). {\it The Theory of Branching Processes}.
Springer, Berlin.

\item
Kantorovich, L.V. and Krylov, V.I. (1958). {\it Approximate Methods of
Higher Analysis}. Interscience Publishers, Inc., New York.

\item
Kent, S. (2000). On the trial of intrusions into information systems. {\it IEEE
Spectrum} {\bf 37}, 52-56.

\item
Kress,~R. (1989). {\it Linear Integral Equations}. Springer, Berlin.

\item
Lorden,~G. (1971). Procedures for reacting to a change in distribution. {\it
Ann.\ Math.\ Statist.} {\bf 42}, 1897-1908.

\item
Lucas,~J.M. (1985). Counted data CUSUM's. {\it
Technometrics} {\bf 27}, 129-144.

\item
Lucas, J.M., and Crosier, R.B. (1982). Fast initial response for
CUSUM quality-control schemes: Give your CUSUM a
head start. {\it Technometrics} {\bf 24}, 199-205.

\item
Lucas, J.M., and Saccucci, M.S. (1990). Exponentially weighted moving
average control schemes: properties and enhancements (with discussion). {\it
Technometrics} {\bf 32}, 1-29.

\item
MacNeill, I.B., and Mao, Y. (1993). Change-point analysis for mortality and
morbidity rate. {\it Journal of Applied Statistical Science} {\bf 1}, 359-377.

\item
Mei, Y. (2006). Comments on: A note on optimal detection of a
change in distribution, by Benjamin Yakir.  {\it Ann. Statist.} {\bf 34},
1570-1576.

\item
Mevorach, Y. and Pollak, M. (1991). A small sample size comparison of the Cusum and Shiryayev-Roberts
approach to changepoint detection. {\it Am. Jour. Math. Manag. Sciences} {\bf 11}, 277-298.

\item
Moustakides,~G.V. (1986). Optimal stopping times for detecting changes in
distributions. {\it Ann.\ Statist.} {\bf 14}, 1379-1387.

\item
Moustakides, G.V. (2004). Optimality of the CUSUM procedure in continuous time. {\it Ann.\
Statist.} {\bf 32}, 302-315.

\item
Moustakides, G.V. (2008). Sequential change detection revisited. {\it Ann.\
Statist.} {\bf 36}, 787-807.

\item
Novikov, A.A. (1990). On the first passage time of an autoregressive process
over a level and an application to a disorder problem. {\it Theory Probab.\
Appl.} {\bf 35}, 269-279.

\item
Novikov, A.A., and Ergashev, B. (1988). Analytical approach to computing of
exponential smoothing algorithm for change-point detection. {\it Statist.\
Control Problems} {\bf 83}, 110-114 (in Russian).

\item
Page,~E.S. (1954). Continuous inspection schemes. {\it Biometrika}
{\bf 41}, 100-115.

\item
Petrovskii,~I.G. (1957). {\it Lectures on the Theory of Integral Equations}.
Graylock Press, Rochester, New-York.

\item
Pollak,~M. (1985). Optimal detection of a change in distribution. {\it Ann.\
Statist.} {\bf 13}, 206-227.

\item
Pollak,~M. (1987). Average run lengths of an optimal method of detecting a
change in distribution. {\it Ann.\ Statist.} {\bf 15}, 749-779.

\item
Pollak, M. and Siegmund, D. (1985). A diffusion process and its applications
to detecting a change in the drift of Brownian motion. {\it Biometrika}
{\bf 72}, 267-280.

\item
Pollak, M. and Tartakovsky, A.G. (2009). Optimality properties of the
Shiryaev-Roberts procedure. {\it Statistica Sinica} {\bf 19}, 1729-1739.

\item
Polunchenko, A. (2009). {\it Quickest Change Detection with Applications to Distributed Multi-Sensor Systems}. Ph.D thesis, University of Southern California.

\item
Polunchenko, A.S. and Tartakovsky, A.G. (2010). On optimality of the Shiryaev-Roberts procedure for detecting a change in distribution. {\em Ann. Statist} (accepted in Nov 2009). Available at ArXiv:0904.3370v2 [math.ST], 25 August 2009.

\item
Quarteroni, A., Sacco, R., and Saleri, F. (2000). {\it Numerical
Mathematics}. Springer, New-York.

\item
Ritov,~Y. (1990). Decision theoretic optimality of the
CUSUM procedure. {\it Ann. Statist.} {\bf 18}, 1466-1469.

\item
Roberts, S.W. (1959). Control chart tests based on geometric moving
averages. {\it Technometrics} {\bf 1}, 239-250.

\item
Roberts, S.W. (1966). A comparison of some control chart procedures. {\it
Technometrics} {\bf 8}, 411-430.

\item
Robinson, P.B., and Ho, T.Y. (1978). Average run lengths of geometric
moving average charts by numerical methods. {\it Technometrics} {\bf 20}, 86-93.

\item
Shiryaev, A.N. (1961). The problem of the most rapid detection of a
disturbance in a stationary process. {\it Soviet Math.\ Dokl.} {\bf 2},
795-799.

\item
Shiryaev, A.N. (1963). On optimum methods in quickest detection
problems. {\it Theory Probab.\ Appl.} {\bf 8}, 22-46.

\item
Shiryaev, A.N. (1996). Minimax optimality of the method of cumulative sum
(CUSUM) in the case of continuous time. {\it Russian Math.\ Surveys} {\bf 51},
750-751.

\item
Srivastava, M.S., and Wu, Y. (1993). Comparison of EWMA,
CUSUM and Shiryaev-Roberts procedures for detecting a shift in the
mean. {\it Ann.\ Statist.} {\bf 21}, 645-670.

\item
Tartakovsky, A.G. (1991). {\it Sequential Methods in the Theory of
Information Systems}. Radio i Svyaz', Moscow (in Russian).

\item
Tartakovsky, A.G. (1995). Asymptotic properties of CUSUM and Shiryaev's
procedures for detecting a change in a nonhomogeneous Gaussian process. {\it
Mathematical Methods of Statistics} {\bf 4}, 389-404.

\item
Tartakovsky, A.G. (2005). Asymptotic performance of a multichart CUSUM test
under false alarm probability constraint. {\it Proc.\ 44th IEEE Conf.\ on Decision and Control and the European Control Conf.\ (CDC-ECC'05)},
320-325. Seville, Spain.

\item
Tartakovsky, A.G. (2009). Discussion on ``Is average run length to false
alarm always an informative criterion?'' by Yajun Mei. {\it Sequential
Analysis} {\bf 27}, 396-405.

\item
Tartakovsky, A.G. and Ivanova, I.A. (1992). Comparison of some sequential
rules for detecting changes in distributions. {\it Probl.\ Inform.\ Transmis.}
{\bf 28}, 117-124.

\item
Tartakovsky, A.G., Li, X.R., and Yaralov, G. (2003). Sequential detection
of targets in multichannel systems. {\it IEEE Trans. Inform. Theory} {\bf 49}, 425-445.

\item
Tartakovsky, A.G., Rozovskii, B.L., Bla\v{z}ek, R., and Kim, H.
(2006). Detection of intrusions in information systems by sequential change-point methods, {\it Statist. Methodology}, {\bf 3}, 252-340.

\item
Tartakovsky,~A.G. and Veeravalli, V.V. (2004). Change-point detection in
multichannel and distributed systems with applications. In {\it Applications of
Sequential Methodologies}, (N. Mukhopadhyay, S. Datta and S.
Chattopadhyay eds.), 331-363. Marcel Dekker, New York.

\item
Wang, H., Zhang, D., and  Shin, K.G. (2002). Detecting SYN flooding
attacks. {\it Proc. INFOCOM'2002, 21st Annual Joint Conference of the IEEE Computer and Communications Societies} {\bf 3}, 1530-1539.

\item
Willsky, A. (1976). A survey of design methods for failure detection in
dynamical systems. {\it Automatica} {\bf 12}, 601-611.

\end{description}

\vskip .65cm
\noindent George V. Moustakides\\
University of Patras, Department of Electrical and Computer Engineering,\\
26500 Rion, Greece
\vskip 2pt
\noindent
E-mail: moustaki@upatras.gr
\vskip 2pt
\noindent
Aleksey S. Polunchenko and Alexander G. Tartakovsky\\
University of Southern California, Department of Mathematics,\\
3620 S. Vermont Avenue, Los Angeles, CA 90089-2532, USA
\vskip 2pt
\noindent
E-mail: (polunche, tartakov)@usc.edu
\vskip .3cm
\end{document}